\documentclass[aps,prl,twocolumn,showpacs,superscriptaddress,groupedaddress]{revtex4}  
\usepackage{graphicx}  
\usepackage{dcolumn}   
\usepackage{bm}        
\usepackage{amssymb}   

\begin{document}



\title{Structural heterogeneity and diffuse scattering in morphotropic lead zirconate-titanate single crystals}

\author{R. G. Burkovsky}\email {rg.burkovsky@mail.ioffe.ru}
\affiliation{St.Petersburg State Politekhnical University, 29 Politekhnicheskaya, 195251, St.-Petersburg, Russia}

\author{Yu. A. Bronwald}
\affiliation{St.Petersburg State Politekhnical University, 29 Politekhnicheskaya, 195251, St.-Petersburg, Russia}

\author{A. V. Filimonov}
\affiliation{St.Petersburg State Politekhnical University, 29 Politekhnicheskaya, 195251, St.-Petersburg, Russia}

\author{A. I. Rudskoy}
\affiliation{St.Petersburg State Politekhnical University, 29 Politekhnicheskaya, 195251, St.-Petersburg, Russia}

\author{D. Chernyshov}
\affiliation{Swiss-Norwegian Beamlines at ESRF, BP 220, F-38043 Grenoble Cedex, France}

\author{A. Bosak}
\affiliation{European Synchrotron Radiation Facility, BP 220, F-38043 Grenoble Cedex, France}

\author{J. Hlinka}
\affiliation{Institute of Physics, Academy of Sciences of the Czech Republic, Na Slovance 2, 182 21 Prague 8, Czech Republic}

\author{X. Long}
\affiliation{Department of Chemistry and 4D LABS, Simon Fraser University, Burnaby, British Columbia, Canada V5A 1S6}

\author{Z.-G. Ye}
\affiliation{Department of Chemistry and 4D LABS, Simon Fraser University, Burnaby, British Columbia, Canada V5A 1S6}

\author{S. B. Vakhrushev}
\affiliation{Ioffe Phys.-Tech. Institute, 26 Politekhnicheskaya, 194021, St.-Petersburg, Russia}
\affiliation{St.Petersburg State Politekhnical University, 29 Politekhnicheskaya, 195251, St.-Petersburg, Russia} 
\date{\today}

\begin{abstract}
Complementary diffuse and inelastic synchrotron X-ray scattering measurements of lead zirconate-titanate single crystals with composition near the morphotropic phase boundary (x=0.475) are reported. In the temperature range 293 K $<$ T $<$ 400 K a highly anisotropic quasielastic diffuse scattering is observed. Above 400 K this scattering disappears. Its main features can be reproduced by model of inhomogeneous lattice deformations caused by inclusions of a tetragonal phase into a rhombohedral or monoclinic phase. This observation supports the idea that PZT at its morphotropic phase boundary is essentially structurally inhomogeneous.
\end{abstract}

\maketitle


Lead zirconate-titanate (PbZr$_{1-x}$Ti$_x$O$_3$, PZT) is one of the most technologically important ferroelectrics \cite{Haertling:1999,Scott:2007}. Being widely employed in practice PZT is also a model system representing ferroelectric solid solutions with a morphotropic phase boundary (MPB). Understanding of the mechanisms leading to very high dielectric and piezoelectric responses near the MPB is essential for strategic design of new and improved materials, particularly the ecologically friendly lead-free PZT counterparts \cite{Cross:2004}. 
Numerous theoretical and experimental studies performed on different lead-containing MPB ferroelectrics produce a highly complex and often controversial picture. Many of these aspects are covered in a 2006 review by Noheda and Cox \cite{Noheda:2006} and references therein. More recent trends are briefly highlighted in a 2009 editorial review by Kreisel et al. \cite{Kreisel:2009}. Recently the morphotropic PZT single crystals of various compositions become available \cite{Bokov:2010}, which has opened up new experimental possibilities. The first single-crystal diffraction experiments \cite{Phelan:2010,Gorfman:2011} did not provide the final conclusion regarding the true microscopic structure of PZT, but allowed to make several important conclusions. On the basis of neutron diffraction study of morphotropic PZT with x=0.46, it was shown \cite{Phelan:2010} that low-temperature monoclinic $Cc$ phase should be ruled out as a ground state and a coexistence of rhombohedral and monoclinic $Cm$ domains should be considered instead. High-resolution X-ray diffraction study \cite{Gorfman:2011} also supports a phase coexistence model for that composition. By using 2-dimensional single-crystal scattering maps instead of 1-dimensional powder spectra the authors succeeded in resolving otherwise overlapping Bragg reflections and demonstrated the presence of more than one phase. The idea of phase coexistence was also supported by recent studies by anelastic and dielectric spectroscopy \cite{Cordero:2011} and neutron powder diffraction \cite{Zhang:2011}.

A number of times it was also pointed out that morphotropic PZT can be inhomogeneous on the nanoscale. This point of view is supported by observation of nanometric contrast fluctuations within micrometer-scale domains in PZT by transmission electron microscopy \cite{Shonau:2007,Asada:2007}. Twinned nanodomains were also considered as a cause for unusual optical isotropy \cite{Bokov:2010} revealed in the tetragonal phase of PZT with x=0.46. From another point of view \cite{Glazer:2004} the nanoscale heterogeneity in PZT can be connected with regions of short-range correlated monoclinic ionic displacements which on average produce the diffraction pattern compatible with rhombohedral and tetragonal symmetry at different sides of the phase diagram. It was also suggested on the basis of single-crystal inelastic X-ray scattering \cite{Hlinka:2011} that morphotropic PZT has similar to relaxors relaxational-type zone-boundary lattice dynamics and shares with them some intrinsic nanoscale inhomogeneity.


\begin{figure*}
\includegraphics [width=1.7\columnwidth,clip=true, trim=0mm 0mm 0mm 0mm] {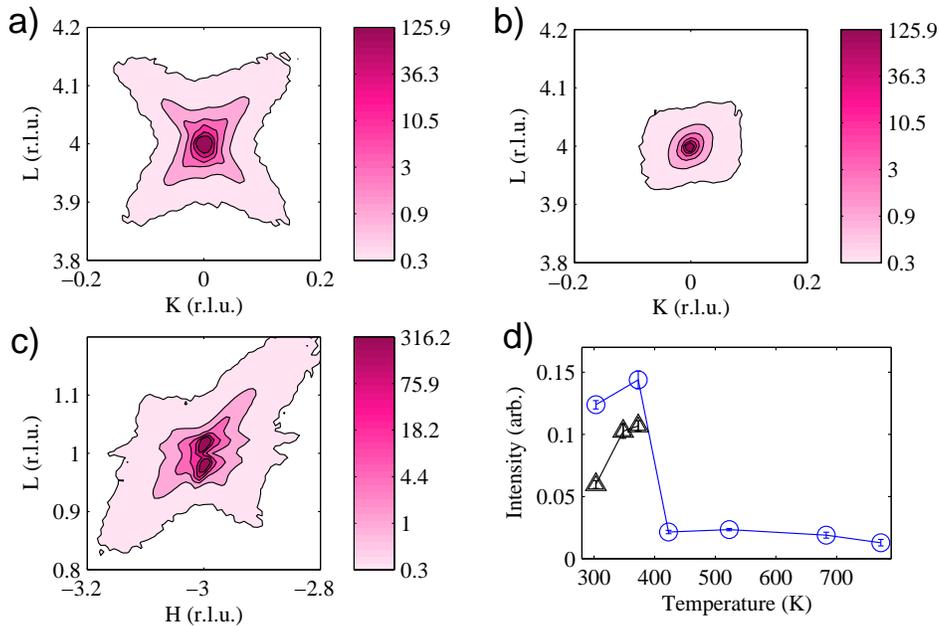}
\caption{\label{fig:2d_maps} Diffuse scattering (DS) distributions in the (0~0~4) zone at room temperature (a) and 683 K (b), DS distribution in the (-3~0~1) zone at room temperature (c), temperature dependence of integrated DS intensity during heating (circles) and cooling (triangles) (d).}
\end{figure*}

A powerful technique of studying structural and other types of inhomogeneities in crystals is diffuse scattering (DS). Particularly it proves itself useful in the studies of structural instabilities in ferroelectrics and related systems \cite{Gehring:2009}. 
To the best of our knowledge, no publications have been available in which the diffuse scattering was observed or interpreted in PZT except the electron diffraction study by Glazer et al \cite{Glazer:2004}. The authors report the observation of DS in both Zr-rich and Ti-rich compositions but not in the ones close to the morphotropic phase boundary. By using single-crystal synchrotron X-ray scattering we show that the diffuse scattering indeed exists in morphotropic composition of PZT, evidencing its structural heterogeneity. 


Morphotropic PbZr$_{1-x}$Ti$_x$O$_3$ single crystals with PbTiO$_3$ content x=0.475 were grown by a top-seeded solution method. For the X-ray measurements a stick-shaped sample with about 100 x 100 micrometer cross-section was prepared by slicing, polishing and subsequent etching in HCl acid. Diffraction and diffuse scattering measurements were carried out at Swiss-Norwegian beamlines at ESRF using KUMA (Oxford Diffraction) diffractometer with CCD detector. A locally constructed heat blower was used for heating up to 773~K. 

At room temperature we obtained the diffuse scattering distributions in the (0~0~4) and (-3~0~1) zones (Figs. \ref{fig:2d_maps}a, and \ref{fig:2d_maps}c). They are highly anisotropic and appear to resemble the DS shapes in relaxors \cite{Xu:2004:Neutron,Bosak:2011}. On heating this strong DS disappears at temperatures between 373~K and 423~K and only weak, but also anisotropic diffuse halo remains up to 773~K. Characteristic distribution of this high-temperature DS is shown in Fig. \ref{fig:2d_maps}b. On cooling the strong DS reappears in the same temperature region, but according to our measurements is systematically less intense than before heating. The temperature dependence of the DS intensity is presented on Fig. \ref{fig:2d_maps}d. The intensity points on that plot correspond to the values of parameter $I_0$ obtained by data fits described below.

\begin{figure} 
\includegraphics [width=.6\columnwidth,clip=true, trim=0mm 0mm 0mm 0mm] {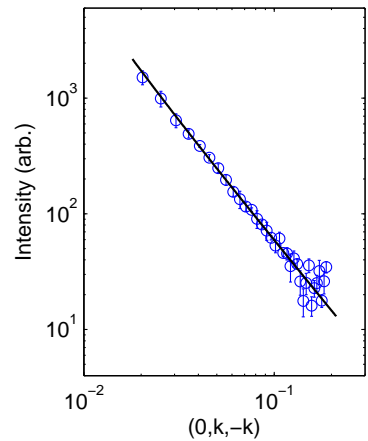}
\caption{\label{fig:profile} Diffuse scattering profile along the [0~1~-1] direction in the (0~0~4) Brillouin zone at room temperature together with the results of fit by expression $I(q) = I_0 q^{-s}$.}
\end{figure}

A log-log plot of diffuse scattering profile along the high-intensity direction [0~1~-1] in the (0~0~4) zone is presented in Fig. 2. The data for $q < 0.02$ r.l.u. are spoiled by Bragg scattering and not included in the plot. For larger $q$ values a power law $I(q)=I_0 q^{-s}$ is observed with the power $s = 2\pm 0.2$. The corresponding fit is shown by solid line. For estimation of the temperature dependence of DS intensity in Fig. \ref{fig:2d_maps}d the value of parameter $s$ was rounded up to be 2.

\begin{figure}
\includegraphics [width=.75\columnwidth,clip=true, trim=0mm 0mm 0mm 0mm] {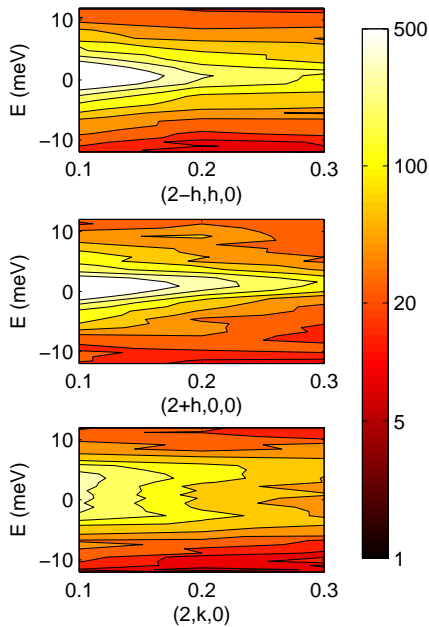}
\caption{\label{fig:inelastic_3} Inelastic X-ray scattering intensity maps for the diagonal, longitudinal and transverse directions in the (2~0~0) Brillouin zone.}
\end{figure}

Diffuse scattering distributions obtained at SNBL include both the elastic and inelastic contributions. To distinguish the nature of this DS we additionally performed an inelastic X-ray scattering (IXS) experiment at ID28 ESRF beamline. The resolution was about 3 meV. The IXS maps along the diagonal (2-h,h,0), longitudinal (2+h,0,0) and transverse (2,k,0) directions are presented in Fig. \ref{fig:inelastic_3}. These maps apparently demonstrate that the maximum signal corresponds to elastic scattering ($E$=0) for diagonal and longitudinal directions but no elastic line is observed for transverse direction. This contrasts with the DS in relaxors where temperature-dependent DS exists in the transverse direction but is almost absent in the longitudinal direction \cite{Vakhrushev:1996}.

Surprisingly, we do not see any increase of X-ray DS intensity near $T$=663~K, where the cubic-to-tetragonal transition takes place. In fact this transition is accompanied by a strong central peak in Brillouin scattering \cite{Kim:2012}, most possibly associated with ferroelectric fluctuations. We do not see such fluctuations by X-rays near high-temperature transition and thus do not expect to observe them near low-temperature transition. This way we see the strong increase of DS intensity below 423~K as a sign of developing heterogeneity. 
A starting point for interpreting this heterogeneity can be set up on the basis of previous results, that indicate the simultaneous presence of multiple phases. 
We start from an assumption that a host phase of specific symmetry contains clusters of a different symmetry. In this case the DS can be described by the terms corresponding to the form factor of the clusters and the factor describing the impact of these clusters on the matrix \cite{Krivoglaz:1996}:

\begin{equation}
I = N_de^{-2W}\frac{1}{v^2\lambda} 
\sum_{\alpha=1}^\lambda
|s_\alpha(\vec{q})|^2
|\Delta f - f\vec{Q}\vec{A}_{\vec{q}\alpha}|^2.
\end{equation}

\begin{figure*}
\includegraphics [width=1.4\columnwidth,clip=true, trim=0mm 0mm 0mm 0mm] {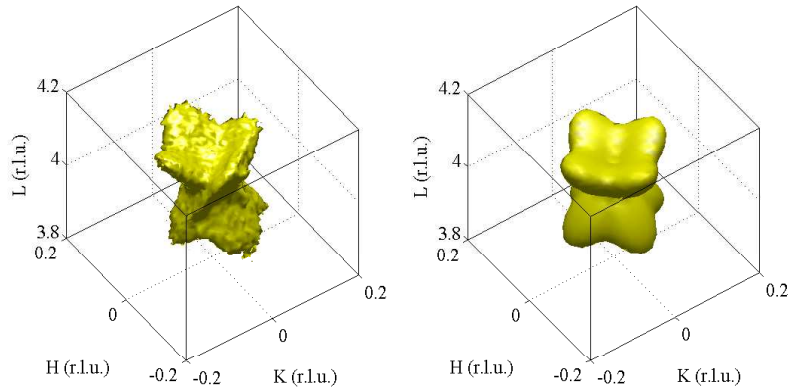}
\caption{\label{fig:3d} A comparison of the experimental 3-D distribution of diffuse scattering (left) and calculations based on the model of Huang scattering due to tetragonal defects (right). Pseudocubic elastic constants c$_{11}=135$, c$_{12}=75$ and c$_{44}=70$ GPa were used for calculations.
}
\end{figure*}

The intensity is proportional to the number of defect centers $N_d$, Debye-Waller factor $e^{-2W}$ and a sum of $\lambda$ additives. Each additive describes the scattering due to particle of orientation $\alpha$. The shape of particle is represented by Fourier transform of corresponding shape function $s_\alpha(\vec{q})$. $\Delta f$ represents the difference between the structure factors of the host phase and the clusters. The term $f\vec{Q}\vec{A}_{\vec{q}\alpha}$ describes the scattering due to elastic deformations caused in matrix by particles. This latter term is assumed to be dominant in comparison with $\Delta f$ since the structures of matrix and particles in PZT are expected to be very close. When particles are sufficiently small we may also neglect a form factor and focus our consideration on elastic deformations. Characteristic 2-D distributions of DS are tabulated in Ref. \onlinecite{Krivoglaz:1996} for many simple defect symmetries and one may easily find that tetragonal defects in cubic matrix will cause DS very similar to the maps on Fig \ref{fig:2d_maps}. We find that the best agreement is achieved when the symmetry of defects is described by characteristic tensor $L$ with only non-zero elements $L_{xx}=L_{yy}=-2L_{zz}$. The defects of this type tend to compress the surrounding lattice in $x$ and $y$ directions while elongating it in the remaining $z$ direction or vice versa. The volume of unit cell tends to be preserved. 
The elastic constants of morphotropic PZT single crystals, needed for calculations, have not yet been determined experimentally, but ceramics data (see Ref. \onlinecite{Ouyang:2006} and references therein) and theoretical estimations \cite{Cohen:2001} are available. Pseudocubic elastic constants extracted from ceramics data \cite{Ouyang:2006} allow us to get a satisfactory qualitative description of DS distributions. However the best agreement we find with slightly changed constants $c_{11}=135$, $c_{12}=75$ and $c_{44}=70$ GPa. The corresponding comparison is presented in Fig. \ref{fig:3d}.

Despite apparent simplicity of our model we find it very reasonable. While we do not know the exact symmetry of the host phase we may assume it to be close to cubic. First because the deviations from cubic structure at least for compositions close to MPB are rather small. Secondly, after averaging over all possible orientations of monoclinic/rhombohedral domains the matrix will appear effectively cubic. The assumption of tetragonal symmetry of clusters is also well-founded since previous studies indicate the signs of tetragonal phase in the MPB region. And, at last, this model perfectly reproduces all the main peculiarities of the DS revealed by our  experiments. It gives a zero-intensity transverse plane in high-symmetry (0 0 L) zones and a non-zero longitudinal scattering. Also it reproduces the observed $q^{-2}$ scattering law. This combination could not be reproduced by purely form-factor based models such as the ones, proposed in Ref. \onlinecite{Xu:2004:Three}.

In summary, in this Letter we reported complementary diffuse and inelastic X-ray scattering measurements on a morphotropic PZT single crystal that in tandem allow us to precisely point to the most probable microscopic organization of the material below the morphotropic phase transition temperature $T_{\textrm{\scriptsize{MPB}}}$. When the temperature falls below $T_{\textrm{\scriptsize{MPB}}}$ the tetragonal phase is not completely destroyed but remains in the form of local inclusions within the host phase of a different symmetry. We do not see any decline in the corresponding DS down to room temperature, which indicates a high stability of the proposed structurally heterogeneous state.

\begin{acknowledgments} 
It is a pleasure to acknowledge P. M. Gehring and A. K. Tagantsev for many useful discussions and various suggestions. 
The work at SPbSPU was supported by Federal Program ``Research and development on high-priority directions of improvement of Russia's scientific and technological complex'' for 2007-2013 years and by grant of the St.-Petersburg government. 
The work at Ioffe institute was supported by RFBR grant 11-02-00687-a.
The work at Institute of Physics was supported by the Czech Science Foundation (GACR P204/10/0616).
The work at SFU was supported by the U.S. Office of Naval Research (Grants No N00014-06-1-0166 and N00014-11-1-0552) and the Natural Sciences and Engineering Research Council of Canada.
\end{acknowledgments}

\bibliography{bibliography_relaxor_pzt}

\end{document}